\documentstyle [12pt]{article}
\newcommand{\n}{\not\!}
\begin{document}
\baselineskip = 21 pt
 \thispagestyle{empty}
 \title{
\vspace*{-2.5cm}
\begin{flushright}
\begin{tabular}{c c}
& {\normalsize MPI-Ph/93-48}\\
& {\normalsize July 1993}
\end{tabular}
\end{flushright}
{}~\\
{}~\\
 The  $\tau$ neutrino as a Majorana particle
 ~\\}
 \author{M. Carena,
 B. Lampe   and C. E. M. Wagner\\
 ~\\
Max Planck Institut f\"{u}r Physik\\
F\"{o}hringer Ring 6\\
D-8000 M\"{u}nchen 40, Germany.\\
{}~\\
{}~\\
 ~\\
 ~\\
 ~\\ }
\date{
\begin{abstract}
A Majorana mass term for the $\tau$ neutrino would induce
neutrino - antineutrino mixing and thereby a process which
violates fermion number by two units.
We study the possibility of distinguishing between a
massive Majorana and a Dirac
$\tau$ neutrino, by
measuring fermion number violating processes in
a deep inelastic scattering experiment $\nu p \rightarrow
\tau X$. We show that, if the neutrino beam is obtained
from the decay of high energetic pions, the probability
of obtaining "wrong sign" $\tau$ leptons is  suppressed
by a factor ${\cal{O}}(m_{\nu_{\tau}}^2 \theta^2/m_{\mu}^2)$
instead of the naively expected suppression factor
$\theta^2 m_{\nu_{\tau}}^2/E_{\nu}^2$, where $E_{\nu}$ is the
$\tau$    neutrino energy,
  $m_{\nu_{\tau}}$ and $m_{\mu}$ are the $\tau$-neutrino and
muon masses, respectively,
and $\theta$ is the $\nu_{\mu}$ -
$\nu_{\tau}$ mixing angle. If $m_{\nu_{\tau}}$ is
of the order of 10 MeV
and $\theta$ is of the order of $0.01 - 0.04$ (the present
bounds are ($m_{\nu_{\tau}} < 35 MeV, \theta < 0.04$) the next
round of experiments may be able to distinguish  between
Majorana and
Dirac $\tau$-neutrinos.
\end{abstract}}
\maketitle
{}~\\
{}~\\
{}~\\
\newpage
\hspace*{-0.6cm}
{\bf{Introduction.}}  Neutrino masses and mixing angles
may provide one of the
simplest test of physics beyond the Standard Model.
Since neutrinos are chargeless, their mass can a priori be
either of Dirac or Majorana type. Experimental measurements
of neutrino masses are, however, very hard and,
in fact, the only hints of nonzero
neutrino masses come from astrophysical processes rather
than from laboratory experiments. Nonzero neutrino masses
provide a way of resolving the  discrepancy between the
predicted and the observed neutrino flux coming from the
sun. Indeed, the so called MSW mechanism\cite{MSW}
relies not only
on nonzero masses but, in analogy to what happens in the
quark sector of the electroweak theory, on nonzero mixing
between the neutrinos of the first two families. If this
mechanism were correct, very precise predictions for the
masses and Cabbibo-like mixing angles of the first two
generations could be obtained from the combination of the
most recent experimental data\cite{Galex}.

A strong constraint on neutrino masses comes from a cosmological
argument\cite{Raffelt}. If all neutrinos were
stable, limits on their masses may be obtained by requiring
the neutrino density to be below the critical density needed
to close the universe, $\sum_{i} m_{\nu_i} < 100$eV.
According to this argument heavy neutrinos with masses in
the range 1 KeV - 100 MeV can only exist
assuming a sufficiently fast
decay rate. Thus,  one has to assume that the same unknown
mechanism
which gives the neutrinos a mass induces suitable decay processes
turning  them unstable.
Moreover, if neutrinos were Dirac particles, limits on
their masses would be also obtained from Supernova events,
$m_{\nu} < 30$ KeV \cite{Kev}.

The bounds on the neutrino masses derived from laboratory
experiments are, instead,
much weaker than the astrophysical bounds. For the
$\tau$ neutrino, for example, the present experimental
limit is $m_{\nu_{\tau}} < 35$ MeV\cite{tauneu}.
This bound comes from an analysis
of final states $X$ in the $\tau$ decay, $\tau \rightarrow
\nu_{\tau} X$.
If the $\tau$ neutrino had a mass
of the order of a few MeV, then
 supernova constraints would prevent it from being a Dirac
particle. If, instead, it is   a Majorana
neutrino, its mass will  not be restricted by
supernova constraints, but it will have to decay sufficiently fast
in order to fulfill the cosmological bound.

In addition,
if our present understanding of nucleosynthesis in the
early universe were correct, a more stringent limit on
the $\tau$ neutrino mass would
be obtained. As a matter of fact, independently of
whether the $\tau$ neutrino is a Dirac or a
Majorana particles,
its  mass would  be bounded
to be $m_{\nu_{\tau}} < 1$ MeV, if its lifetime were
larger than ${\cal{O}}$(100 sec.). On the contrary, if
its lifetime were shorter or of the order of
10 sec., no bound on its mass would appear
from these considerations \cite{Steigman}.
It is easy to show that this requirement
would not be fulfilled if the only physics beyond the standard
model were  the neutrino masses and mixing angles and
hence a $\tau$ neutrino mass in the MeV range would imply either a
revision of our present understanding of nucleosynthesis, or
exciting new physics in the TeV range.

In the following, we shall consider $\tau$-neutrinos with
Majorana masses in the MeV range, produced in a deep
inelastic experiment. Thus, we should also assume that
the same mechanism which gives the $\tau$-neutrino a mass makes
it unstable so that cosmological and astrophysical constraints
do not apply. Furthermore, one should keep in mind that it is
always important to crosscheck astrophysical predictions with
laboratory experiments.

\hspace*{-0.6cm}
{\bf{Majorana masses and mixing.}}
As we just mentioned, for the extent of this work we shall
consider the neutrinos to be massive Majorana particles.
Therefore, neutrino-antineutrino mixing becomes possible.
In fact, whereas a Dirac mass term
\begin{equation}
{\cal{L}}_D = m_D \bar{\nu}_L \nu_R + h.c.
\end{equation}
flips the spin of the neutrinos, a Majorana mass term
\begin{equation}
{\cal{L}}_M = m_M \nu_L^T C \nu_L + h.c.
\end{equation}
mixes neutrinos with antineutrinos, where $C$  is the
charge conjugation matrix and $\nu_{L,R}$ are the left
handed and right handed components of the neutrino
field, respectively. This is possible due to the fact that a Majorana
field is invariant under CP conjugation so that transitions
between particle and antiparticle are allowed, with
probability $\simeq  m_M^2$. If one has more than
two Majorana neutrinos, e.g. $\nu_{\mu}$ and $\nu_{\tau}$,
then there is in general a mass matrix M,
\begin{equation}
{\cal{L}}_M =
( \nu_{\mu}^T, \nu_{\tau}^T)
 C M \left( \begin{array}{c} \nu_{\mu} \\ \nu_{\tau}
\end{array}
\right)
\end{equation}
and mass
eigenstates $\nu_1$, $\nu_2$ should be introduced
\begin{equation}
\left(\begin{array}{c} \nu_{\mu} \\ \nu_{\tau} \end{array}
\right)
= \left(
\begin{array}{cc} \cos\theta  & - \exp(- i \delta) \sin\theta \\
\exp(i \delta) \sin\theta   & \cos\theta \end{array}
\right) \left( \begin{array}{c} \nu_1 \\ \nu_2 \end{array}
\right).
\end{equation}
Note that, in the above, contrary to the $2 \times 2$
Dirac type mixing matrix\cite{CPviol}, in
addition to the mixing angle $\theta$ a phase $\delta$ appears.

In the case of a free neutrino beam,
one can show, based on simple dimensional arguments, that
neutrino - antineutrino oscillations will be suppressed in comparison
with the usual neutrino - neutrino oscillations by a factor
$(m_{\nu_{\tau}}/E_{\nu})^2$, where  $E_{\nu}$ is the
neutrino energy,
and hence, for highly energetic neutrinos,
these oscillations will be unobservably small \cite{oscill}.
However, neutrino-antineutrino
oscillations may be experimentally observed if the neutrinos
proceed from a highly energetic pion beam. The physics behind
this observation is extremely simple and is based on the
helicity properties of the neutrinos as observed from the
laboratory frame \cite{Stodolsky}.
Indeed, if one considers
the frame in which the pion is at rest, the neutrinos will
be emitted with an energy
$E^* = (m_{\pi}^2 - m_{\mu}^2)/2m_{\pi}\simeq 30$ MeV, where
$m_{\pi}$ and $m_{\mu}$ are the pion and muon masses, respectively,
and with
definite helicity. When observed from the laboratory
frame, however,
a fraction of them, namely those which were moving in
the backward direction with respect to the beam in the rest frame,
will appear to have opposite helicity and, hence, they behave like
Majorana antineutrinos. In Ref.\cite{Stodolsky}, this fraction
was estimated to be of the order
\begin{equation}
\epsilon \simeq \left(\frac{m_{\nu}}{E^*}\right)^2
\end{equation}
If only the $\mu$ neutrino oscillations are studied,
since due to
experimental limits  its mass can not be larger than a few
tens of keV, then, as observed in Ref.\cite{Stodolsky},
this fraction is unobservably small.
If the  $\mu$ neutrino
has a small admixture with a massive $\tau$ neutrino,
with mass in the MeV range, this
process, although suppressed by mixing angles, can  be
significantly enhanced. In this letter, it is our intention
to make a detailed analysis of deep inelastic scattering
experiments, $\nu p \rightarrow \tau X$, in which the $\nu$
beam is obtained from high energetic pions. We shall  show
how, in the light of
future  neutrino oscillation experiments, such analysis
may provide  very useful information about the nature of
massive neutrinos and their masses.

\hspace*{-0.6cm}
{\bf{The Experiment.}}
The present experimental search for $\nu_{\mu}$ -
$\nu_{\tau}$ mixing \cite{exper}
is based on a very simple physical
principle: If a high energy beam of charged pions
is produced in the laboratory, the $\pi^+$ ($\pi^-$)
will rapidly decay into a $\mu^+$ ($\mu^-$) and a
$\nu_{\mu}$ ($\bar{\nu}_{\mu}$).  If the neutrino beam
subsequently collides with an isoscalar target, $\mu^-$
($\mu^+$) particles will be produced, which may be easily
detected experimentally. Moreover, considering that
 the electroweak
eigenstate $\nu_{\mu}$ is an admixture of two massive
particles $\nu_1$ and $\nu_2$, with $\nu_2$ being
predominantly a $\nu_{\tau}$ particle, there will be
a nonzero probability of producing $\tau^-$
($\tau^+$) after the collision. The detection of $\tau$
particles in the final states would provide  the
first experimental clear signature of masses and mixing
angles in the neutrino sector, and hence of physics
beyond the standard model.

Experiments like this have been carried out at Fermilab
and new experiments are designed at CERN and Fermilab \cite{proposal}.
The typical experimental setup is as follows: Assume
the pions of energy ${\cal{O}}(100$ GeV) move along the z axis
in the direction of the target. After the pion decay pipe
(of order $\leq$ 1 km), the neutrinos which are emitted
mainly in the forward direction may
oscillate along an oscillation pipe of length ${\cal{O}}
(1$ km) before they hit the target, which has a typical
extension ${\cal{O}}$(1-10 $m^2$). For neutrino masses
in the MeV range, the oscillation is too fast to be observable,
so  the exact length of the
oscillation pipe is not relevant for our analysis. According
to the experimental setup one may introduce cuts  on the angle
$\vartheta$ between the pion and neutrino beam,
$\vartheta < \vartheta_{cut}$, so that
the neutrino beam hits the target,
and on the $\tau$ lepton energy,
 $E_{\tau} > E_{cut}$, so
that $\tau$ lepton identification is guaranteed. Furthermore,
as it is usual in deep inelastic scattering experiments, a cut
$Q^2 > Q_{cut}^2$ with $Q_{cut} = {\cal{O}}$(1 GeV), is introduced to
ensure the validity of the quark parton model involved in
the analysis.

The neutrino oscillation and mixing experiment under consideration
leads in general to excluded areas in the $\theta - \Delta m$ plane,
where $\Delta m = m_2 - m_1$ is the difference between the
mass eigenvalues of the two neutrinos. For our case
($m_2 \simeq {\cal{O}}$(1 MeV), $m_1 \simeq 0$) the oscillations
are too fast, and consequently a limit $\theta \leq 0.04$ can be
deduced from the non observation of $\tau$ leptons in the Fermilab
data\cite{exp}.
 The proposed two new experiments are estimated to increase
the experimental accuracy by at least a factor of ten. Therefore,
it is conceivably that $\tau$ events will be
 observed, a small
fraction of which will
violate lepton number conservation, so that the nature
of the corresponding $\tau$ neutrino mass term is revealed.
The diagrams which describe the reactions are shown in Fig.1.
Also shown in Fig. 1
are the corresponding reactions with a
$\nu - \bar{\nu}$ mixing Majorana propagator which leads to
$\tau$ - leptons of opposite charge.

\hspace*{-0.6cm}
{\bf{Amplitudes and Cross Sections.}}
The total cross section consists on three different processes,
the pion decay, the neutrino oscillations and finally the
scattering process. Within a good approximation,
for the case of $\nu_{\mu}  -
\nu_{\tau}$ ($\bar{\nu}_{\mu} - \bar{\nu}_{\tau}$)
oscillations the total cross section
factorizes, that is to say, it is a product of
the pion decay rate, the oscillation factor and the deep
inelastic lepton scattering. In the case of neutrino -
antineutrino oscillations, the process is a little more
involved and the above factorization does not hold. Following
Ref.\cite{Wilczek}, we compute the total amplitude as follows.
We first consider the $\pi^+$ decay amplitude
to be given by
\begin{equation}
M_{i\mu} =  \frac{G_F}{\sqrt{2}} f_{\pi} \bar{u}_i(k_i)
\left[ m_i (1 - \gamma_5) - m_{\mu}(1 + \gamma_5)
\right] v_{\mu}(k_{\mu})
\end{equation}
were $G_F$ and $f_{\pi}$ are the Fermi and pion decay constants,
respectively, $m_i$ is the mass of the  neutrino mass eigenstates
$\nu_i$ and $k_i$, $k_{\mu}$ and $k_{\tau}$ are the four momenta of the
neutrino $\nu_i$, the $\mu$ and     $\tau$ leptons, respectively.
The form of the $\pi\mu\nu$ interaction vertex is
essentially dictated by the pseudoscalar nature of the pion.
Analogously,
for the $\pi^-$, its decay amplitude is given by
\begin{equation}
M_{i\mu} = \frac{G_F}{\sqrt{2}} f_{\pi} \bar{u}_i(k_{i})
\left[ m_i (1 + \gamma_5) - m_{\mu}(1 - \gamma_5)
\right] v_{\mu}(k_{\mu}).
\end{equation}
The amplitude for the scattering process of the neutrino
against an isoscalar target, with a $\tau^{-}$ in the final
state, reads,
\begin{equation}
T_{\nu_i-N} =
\frac{G_F}{\sqrt{2}} <J_{\nu}>
\bar{u}(k_{\tau}) \gamma^{\nu}(1 - \gamma_5)
u_i(k_i)
\end{equation}
where $J_{\nu}$ is the hadronic charged current amplitude.
For an antineutrino, the corresponding scattering process
leads to
\begin{equation}
T_{\bar{\nu}_i-N} =
\frac{G_F}{\sqrt{2}} <J_{\nu}>
\bar{v}(k_{\tau}) \gamma^{\nu}(1 - \gamma_5)
v_i(k_i)
\end{equation}
The total process amplitude is a product of these two
amplitudes multiplied by the factors associated to the
mixing angles and the neutrino time evolution factors.
Therefore, the total amplitude for  $\pi^+
\rightarrow \mu^+ \tau^- X$ is given by
\begin{equation}
T_{\pi^+ \rightarrow \tau^-} = \sum_i
M_{i\mu} T_{\nu_i N} U_{i\mu} U_{i \tau}^*
\exp( -i E_i t),
\end{equation}
where $U_{ij}$ are the elements of the $\nu_{\mu}$ -
$\nu_{\tau}$ mixing mass matrix,
\begin{equation}
U = \left[ \begin{array}{clcr}
 \cos\theta & -\sin\theta
\exp(-i \delta) \\
\sin\theta \exp(i \delta) & \;\;\; \cos\theta
\end{array}
\right] .
\end{equation}
For the lepton number
violating case, instead, the total amplitude
reads
\begin{equation}
T_{\pi^+ \rightarrow \tau^+} = \sum_i
M_{i\mu} T_{\bar{\nu}_i N} U_{i\mu} U_{i \tau}
\exp( -i E_i t).
\end{equation}
The generalization for the $\pi^-$ case is straightforward.

{}From the above expressions it is easy to get the total
cross sections for the processes by computing the square of
the probability amplitude. In order to do
this the hadronic matrix elements of the charged currents
should be given. These
matrix elements may be parametrized by \cite{Nacht}
\begin{eqnarray}
W^{(\nu \bar{\nu})}_{\lambda \rho}(p,q) & = &\sum_X
\frac{1}{2M}(2\pi)^3 \delta(p' - p - q)
\sum_{Spins} <N(p)|J_{\lambda}^{\pm}|X(p')>
<X(p')|J_{\rho}^{\pm}|N(p)>
\nonumber\\
& = & \left( - g_{\lambda \rho} + \frac{q_{\lambda} q_{\rho}}
{q^2} \right) W_1^{(\nu,\bar{\nu})}(\nu,Q^2)
- \frac{i}{2M^2}
\epsilon_{\lambda\rho\alpha\beta} p^{\alpha} q^{\beta}
W_3^{(\nu,\bar{\nu})}(\nu,Q^2)
\nonumber\\
& + &
\left(p_{\lambda} - \frac{(pq) q_{\lambda}}{q^2} \right)
\left(p_{\rho} - \frac{(pq) q_{\rho}}{q^2} \right)
\frac{1}{M^2}
W_2^{(\nu,\bar{\nu})}(\nu,Q^2),
\end{eqnarray}
where $q = k_{\nu} - k_{\tau}$, $p$ is the nucleon momenta
and $\nu = pq/M$, with $M$ the nucleon mass. We treat the W exchange
in the Fermi limit, i.e. we assume that $Q^2=-q^2$ is much smaller
than $m_W^2$.This assumption is already implicit in eqs. 6 to 9.
We have only considered the dominant structure functions,
which are related to the scaling functions $F_i(x)$
by
\begin{equation}
F_1(x) = 2 M W_1(\nu,Q^2) \;\;\;\;\;\;\;\;
F_2(x) = \nu W_2(\nu,Q^2) \;\;\;\;\;\;\;\;
F_3(x) = \nu W_3(\nu,Q^2),
\end{equation}
where $x = Q^2/2M\nu$.
In the parton model, the structure functions $F_i(x)$ may
be expressed
 as functions of the quark densities. In the following, in
 order to
make quantitative estimates we have used the parametrization
given in ref. \cite{MT}.

One important consequence of the hadronic current structure is
that the neutrino scattering process is enhanced in comparison
to the antineutrino one. In fact, this can be observed, for
example, in the approximation in which
the charged lepton and neutrino masses are neglected, and ignoring
small scaling violation terms appearing in the antineutrino
cross section. The differential cross section for neutrino
scattering against an isoscalar target is then
given by
\begin{equation}
\frac{\partial \sigma^{(\nu N)}}{\partial x \partial y} =
\frac{G_F^2 M E_{\nu}}{\pi} \left( q(x) +
\bar{q}(x) (1 - y)^2 \right),
\end{equation}
while for an antineutrino it reads,
\begin{equation}
\frac{\partial \sigma^{(\bar{\nu} N)}}{\partial x \partial y} =
\frac{G_F^2 M E_{\nu}}{\pi} \left( q(x) (1 - y)^2 +
\bar{q}(x)  \right),
\end{equation}
where $q(x) = x [N_u(x) + N_d(x)]$,
$\bar{q}(x) = x [N_{\bar{u}}(x) + N_{\bar{d}}(x)]$,  and
$N_{u(\bar{u})}$ and $N_{d(\bar{d})}$ are the up  and down
quark (antiquark) densities of the proton, respectively, and
$y = \nu/E_{\nu}$  is the fraction of neutrino energy transfered
to the charged lepton. It is then
obvious to observe, that since
$q(x) \gg \bar{q}(x)$, the total neutrino cross section is
enhanced by approximately
a factor three with respect to the antineutrino one.

Apart from the hadronic charged current matrix elements the
neutrino oscillation process is important. For the lepton number
conserving processes (l.c.)
the relevant mass and mixing angle
dependent factor is given by
\begin{equation}
F_{osc.}^{l.c.}
 = \sum_{ij} U_{i\mu} U^*_{i\tau}
U^*_{j\mu} U_{j \tau} \exp\left[ i \left(E_j - E_i
\right) t \right]
\end{equation}
which can be approximately expressed as
\begin{equation}
F_{osc}^{l.c.} = 2 \cos^2\theta \sin^2\theta
\left[ 1 - \cos\left( \frac{ \Delta m^2 t }{2 E_{\nu}}  \right)
\right],
\end{equation}
where the second term between brackets gives
the usual oscillation effect. For
one of the masses in the MeV range, the oscillations in any
of the planned experiments would be so fast that, once the
average over the pion decay position is performed,
the oscillation
term gives no contribution to the total cross section.

For the lepton number violating process (l.v.),
 instead, the oscillation
term depends quadratically on the neutrino masses. It is given
by
\begin{eqnarray}
F_{osc}^{l.v.} & = & \sum_{i j} U^*_{i\mu} U_{j \mu}
U^*_{i \tau} U_{j \tau} m_i m_j \exp\left[ i
\left( E_i - E_j \right) t \right]
\nonumber\\
& = & \sin^2 \theta \cos^2 \theta \left[ m_2^2 + m_1^2 -
2 m_1 m_2 \cos\left(\frac{\Delta m^2}{2 E_{\nu}} t +
2 \delta \right) \right].
\end{eqnarray}
Observe that, in this case the cross section has a nontrivial
dependence on the CP violating parameter $\delta$. This is
a very relevant property of Majorana neutrinos, for which,
in contrast to the Dirac neutrino cases, there is no need of
mixing of the three generations to
obtain CP violating effects.
Unfortunately, it can be easily seen that, for neutrino mass
values consistent with the present experimental constraints,
the CP violating contribution is unobservably small and only
the first term $m_2^2$ in the above
expression gives a relevant
contribution ($m_2 \gg m_1$).

\hspace*{-0.6cm}
{\bf{Results.}}
The total process can now be straightforwardly computed. We
shall not give the details here, but just refer to the
relevant properties. The ratio of the lepton number
conserving to the lepton number violating process rate, in the
case of
initial $\pi^{\pm}$ state is given by
\begin{equation}
\frac{\Gamma^{l.v.}}{\Gamma^{l.c.}} =
\frac{m_2^2}{2} \frac{\int\; V_{\alpha \alpha'} W_{\alpha\alpha'}}{
\int\; C_{\alpha \alpha'} W_{\alpha \alpha'}}
\label{eq:20}
\end{equation}
where the first factor comes from the ratio of the oscillation
factors, and the matrix elements $V_{\alpha \alpha'}$ and
$C_{\alpha \alpha'}$ are given by
\begin{eqnarray}
V_{\alpha \alpha'} & = & 2 \left(m_{\pi}^2 + m_{\mu}^2\right)
Tr\left[ \left( 1 \mp \gamma_5 \right)\n k_{\nu}
\gamma^{\alpha} \n k_{\tau} \gamma^{\alpha'}   \right]
\nonumber\\
& + &
2 m_{\mu}^2 Tr\left[ \left( 1 \mp \gamma_5 \right)
\n k_{\mu}
\gamma^{\alpha} \n k_{\tau} \gamma^{\alpha'}  \right]
\end{eqnarray}
\begin{equation}
C_{\alpha \alpha'} = 2 m_{\mu}^2 \left( m_{\pi}^2 -
m_{\mu}^2 \right)
Tr \left[[ \left( 1 \pm \gamma_5
\right) \n k_{\nu} \gamma^{\alpha}
\n k_{\tau} \gamma^{\alpha'} \right] ,
\end{equation}
and an integration over phase space is understood.
In order to get an idea of the order of magnitude of the
above ratio, Eq.(\ref{eq:20}),
 we can use the
approximation $k_{\nu} \simeq k_{\mu}$.
In this case, it is easy to show that,
approximately ($m_2 \simeq m_{\nu_{\tau}}$)
\begin{equation}
\frac{\Gamma^{l.v.}}{\Gamma^{l.c.}} \simeq \frac{m_{\nu_{\tau}}^2}
{2 m_{\mu}^2}
\frac{ \left(
m_{\pi}^2 + 2 m_{\mu}^2 \right) }{
\left( m_{\pi}^2 - m_{\mu}^2 \right)}
\frac{ \sigma^{\nu(\bar{\nu})}}{\sigma^{\bar{\nu}(\nu)}}
\end{equation}
Observe that, as we anticipated, the lepton number
violating process is suppressed with respect to the lepton
number conserving one by a factor of order of the ratio
of the massive neutrino mass to the muon mass squared.
Indeed, if the above approximation were correct, and
based on the approximate magnitude of the $\nu$ and
$\bar{\nu}$ scattering cross section we would get approximately
a 6 $\%$ effect in the case of having a
$\pi^-$ in the initial state, and a
0.7 $\%$ effect in the initial
$\pi^+$ case, when choosing
$m_{\nu_{\tau}} = 10$ MeV.  The actual numbers for the above
ratio, Eq.(\ref{eq:20}), as we shall show below, are
somewhat larger.

In Figures 2, 3 and 4 we present the resulting
ratios of cross section for different experimental cuts and initial
pion energies. We observe
that the percentage of lepton number
violating events is roughly 1 $\%$ for $\pi^+ N$ and
$8 \%$ for $\pi^- N$ scattering, rather independent of the value
of $E_{\pi}$ (see Fig.2)
and $E_{\tau_{cut}}$ (see Fig. 3).
{}From Fig. 2 it also follows that
the ratio of the lepton number violating to the lepton
number conserving process
increases slightly if one
relaxes the cut on the angle between the pion and neutrino
beams,
$\vartheta_{cut}$, while Fig. 4 shows that it depends somewhat
stronger on the value of $Q^2_{cut}$.

\hspace*{-0.6cm}
{\bf{Conclusions.}}
In this letter we have worked out the following idea: If the
neutrinos have masses, the $\tau$-neutrino
is probably the one with
the largest mass. A priori, this can be either of Dirac or of
Majorana type. We have discussed an experiment which
distinguishes between a Majorana and a Dirac mass term.
The main
 problem is that,  if obtained from $\tau$ decays, $\tau$
neutrinos are high energetic and  a neutrino mass determination
with high energy neutrinos is difficult since mass effects are
in general suppressed by powers of $m_{\nu_{\tau}}^2/E_{\nu}^2$.
 Therefore,
we chose the process $\pi N \rightarrow \tau X$, where instead
of a suppression proportional to the inversed $\tau$-neutrino
energy squared,
we found a suppression proportional to the inverse of the muon
mass squared. This is
much less severe and allows to determine  Majorana neutrino
masses in the MeV range. The price we have to pay for this
advantage is an assumption about $\nu_{\mu} - \nu_{\tau}$
mixing which is needed for the process to take place, so that a
suppression factor $\theta^2$ appears throughout in the cross
section. For neutrino masses of the order of a few MeV, the
results look promising. The lepton number violation process
rate $\pi^{\pm}N \rightarrow \mu^{\pm} \tau^{\pm} X$ is of
the order of a few percent of the lepton number conserving
$\pi^{\pm}N \rightarrow \mu^{\pm} \tau^{\mp} X^{'}$ one, the
exact percentage depending on the $\tau$-neutrino mass and
the charge of the initial pion state.
We have also commented on astrophysical
and cosmological restrictions. The essential point is that
cosmological arguments exclude stable neutrinos of
mass ${\cal{O}}(MeV)$. This is due to the fact that
they would contribute
too much to the present mass energy density of the universe.
We have ignored this bound, together with the constraints on
neutrino masses coming from our present understanding of
nucleosynthesis,  by assuming
that the same mechanism which gives the $\nu_{\tau}$ a Majorana
mass makes the $\tau$ neutrino unstable and with a lifetime
such that these bounds do not apply. \\
{}~\\
{}~\\
ACKNOWLEDGEMENTS. The authors would like to thank L. Stodolsky,
G. Raffelt and R. Peccei for interesting conversations.


\end{document}